\documentstyle[sprocl]{article}

\input boxedeps.tex     
\SetOzTeXEPSFSpecial   
\HideDisplacementBoxes

\def\Journal#1#2#3#4{{#1} {\bf #2}, #3 (#4)}

\def\JMP{\em J. Math. Phys.}

\def\PRL{\em Phys. Rev. Lett.}
\def\PRB{{\em Phys. Rev.} B}

\def\JPCS{\em J.~Phys.~Chem.~Solids}

\def\be{\begin{equation}}
\def\ee{\end{equation}}
\def\bea{\begin{eqnarray}}
\def\eea{\end{eqnarray}}
\def\vec#1{\overrightarrow{#1}}

\begin{document}

\title{HIGHER SYMMETRIES IN CONDENSED MATTER PHYSICS}

\author{R. S. MARKIEWICZ, M. T. VAUGHN}

\address{Physics Department, Northeastern University\\
Boston, MA 02115, USA\\E-mail: markiewic@neu.edu, mtvaughn@neu.edu} 


\maketitle\abstracts{We discuss some applications of higher symmetry groups
to condensed matter systems. We give special attention to the groups 
$SO(n)$ $(n = 4, 5, 6, 8)$ in the two-dimensional Hubbard model and its
generalizations, which model the high $T_c$ cuprate superconductors.}

This talk is intended to be a brief introduction to the interesting roles
played by higher symmetries in the theory of the cuprate high $T_c$
superconductors and other quasi-two-dimensional condensed matter systems.
Since this in mainly a particle physics audience, I will give a quick
introduction to the systems we are dealing with. A broad modern survey 
of superconductivity is given by Tinkham$\>$\cite{tink}, for example.

\section{High $T_c$ superconductors}\label{sec:highTc}

\begin{figure}[tb]
\centerline{
\BoxedEPSF{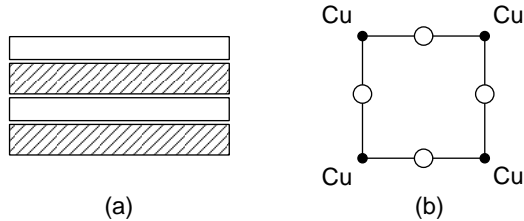 scaled 500}  
}
\vskip 6pt
\noindent
\caption{Schematic views of the cuprate high $T_c$ superconductors.
The left figure (a) shows the conducting $CuO_2$ layers (white) between the 
cross-hatched layers which serve as charge reservoirs, as explained in the text. 
The right figure (b) shows a typical $CuO_2$ layer, with $Cu$ atoms (dark circles) 
at the vertices of a square lattice and $O$ atoms (open circles) at the midpoints 
of the lines joining adjacent $Cu$ atoms.}
\label{fig:CuO2}
\end{figure}

The typical cuprate superconductor has $CuO_2$ layers sandwiched 
between various layers, which serve as charge reservoirs to
provide carriers (holes, in fact) to the conducting $CuO_2$ planes, as
shown in fig.~\ref{fig:CuO2}. A typical charge reservoir is
$La_{2-x} Sr_x O_2$, in which replacing $La^{3+}$ by $Sr^{2+}$ leads
to a hole density $x$ in the conducting layers. High $T_c$ is
typically obtained for doping fractions $x\sim 0.15 - 0.2$. Since
superconductivity is mainly in the $CuO_2$ planes, it is useful
to treat the superconducting system as two-dimensional, with effects
due to coupling between the layers ignored at first.

\section{Two-dimensional lattice}\label{sec:twodlat}

A realistic model Hamiltonian for the $CuO_2$ planes involves electrons (or holes, 
as is the case in practice) hopping between $p$-shell orbitals in $O$ and $d$-shell 
orbitals in $Cu$ together with interaction terms which model the Coulomb repulsion 
between two electrons on the same atom, or on adjacent $Cu$ and $O$ atoms. Detailed 
models are reviewed by Kresin, Morawitz and Wolf$\>$\cite{KMW}. However, it has been
suggested$\>$\cite{Rice} that much of the essential physics of the $CuO_2$ planes is 
captured by a simpler model in which holes only hop between $Cu$ sites; a more detailed
exposition of this view is given in the recent work of Anderson$\>$\cite{PWA}.

\begin{figure}[tb]
\centerline{
\BoxedEPSF{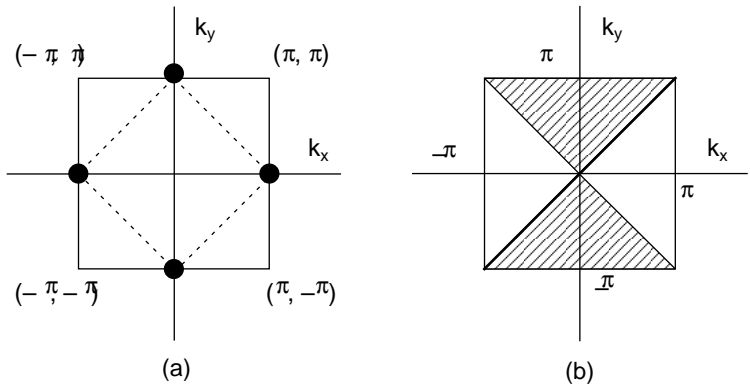 scaled 500}  
}
\vskip 6pt
\noindent
\caption{First Brillouin zone for the square lattice. (a) The dashed lines show the
Fermi surface for the simplest nearest neighbor hopping Hamiltonian (\ref{eq:Hhop})
with $t^{\prime} = 0$ and one electron per site ({\em half filling} of the band).  
The large circles at  ${\bf k} = (0,\pm\pi)$, $(\pm\pi,0)$  give the location of the 
Van Hove singularities in the density of states (see text). (b) Division into two
subzones $\cal X$ and $\cal Y$: $\cal X$ contains the unshaded region together with
the line $k_x + k_y = 0$ and $\cal Y$ contains the shaded region and the dark line
$k_x = k_y$.}
\label{fig:SQlattice}
\end{figure}

Thus we consider first the hopping Hamiltonian
\begin{equation}
H_0 = -\>t\!\sum_{<{\bf xy}>,\alpha}\>a^{\dag}_{\bf x\alpha}a^{\vphantom\dag}_{\bf y\alpha} 
- t^{\prime}\!\!\sum_{<\!\!<{\bf xy}>\!\!>,\alpha}\>a^{\dag}_{\bf x\alpha}
a^{\vphantom\dag}_{\bf y\alpha} 
\label{eq:Hhop}
\end{equation}
Here $<\ >$ ($<\!\!<\ >\!\!>$) denotes sum over (next) nearest neighbor pairs, and
$\alpha$ is a spin index. The spectrum of $H_0$ consists of a single energy band, 
with energies
\begin{equation}
E_{\bf k} = -2t(\cos k_x + \cos k_y) - 4t^{\prime} \cos k_x\cos k_y
\label{eq:Eband}
\end{equation}
as ${\bf k}$ runs over the first Brillouin zone, which is depicted in fig.~\ref{fig:SQlattice}(a).
At the points indicated by large circles in the figure, we have $\nabla_{\bf k} E = 0$,
and hence a singularity in the density of states, the {\em Van Hove singularity} (VHS), which
is logarithmic in two dimensions. One of us (RSM) has written an extensive review$\>$\cite{vHs}
describing the importance of the VHS in high $T_c$ physics.

Another feature important in understanding this lattice is the existence of a 
{\em nesting vector} ${\bf Q}$ such that $\exp(i{\bf Q\cdot x}) = \pm 1$ for every 
lattice site ${\bf x}$. This vector connects the Van Hove singularities, and it also joins 
points near the Fermi surface for electron (or hole) densities near one per site. 
This creates the possibility of various instabilities associated with wave vector {\bf Q},
including antiferromagnetism, d-wave superconductivity, and others. Here the vector
${\bf Q} = (\pm\pi, \pm\pi)$ when length is expressed in units of the lattice spacing.

To describe the interactions of the electrons, we can start with the Hubbard
model$\>$\cite{Hubbard} interaction
\begin{equation}
V_{Hubbard} = U\,\sum_{\bf x}\>n_{\bf x \uparrow}n_{\bf x \downarrow}
\label{eq:UHub}
\end{equation}
which represents the Coulomb interaction between two electrons on the same site.
Schulz$\>$\cite{Schulz} has considered additional interaction terms.

\section{Pairing algebras}\label{sec:pairing}
Consider fermion creation and annihilation operators $a_i^{\dag}, a_i$
($i = 1,\ldots, N$). The $N^2$ number conserving operators $a_i^{\dag}a_j$ generate
an $SU(N)\oplus U(1)$ Lie algebra, and when we combine them with the $N(N-1)$ pairing
operators
\begin{equation}
\Pi_{jk}\equiv a^{\vphantom\dag}_i a^{\vphantom\dag}_j\hbox{\ \ \ and\ \ \ }
\Pi^{\dag}_{jk}\equiv a_j^{\dag}a_i^{\dag}
\label{eq:pijk}
\end{equation}
we obtain a {\em pairing algebra} which generates a pairing group $SO(2N)$. Some elements
of this algebra may be conserved; other elements $\cal O$ may satisfy a commutation relation
of the form
\begin{equation}
[H, {\cal O}] = \lambda {\cal O}
\label{eq:SGA}
\end{equation}
with the Hamiltonian $H$. These elements define a {\it spectrum generating algebra}, since
for an operator $\cal O$ which satisfies (\ref{eq:SGA}),
\begin{equation}
H\psi = E\psi \Rightarrow H({\cal O}\psi) = (E + \lambda)({\cal O}\psi)
\label{eq:Hladder}
\end{equation}
Thus the operator $\cal O$ transforms one energy eigenstate into another with eigenvalue
different by $\lambda$. The first systematic study of these algebras in condensed matter 
physics appears to have been done by Birman and Solomon$\>$\cite{BirSol}.

\section{A simplified SO(8) algebra}\label{sec:SimpleSO}

A simplified $SO(8)$ algebra can be constructed if we let $a^{\dag}_{\alpha}$, 
$a^{\vphantom\dag}_{\beta}$ be creation and annihilation operators for a fermion with momentum 
near $(\pm\pi, 0)$ ($\alpha$, $\beta$ are now spin indices), and $b^{\dag}_{\alpha}$, 
$b^{\vphantom\dag}_{\beta}$ the corresponding operators for momentum near $(0, \pm\pi)$. 
This choice is motivated by the existence of the Van Hove singularities in the density of states, 
which lead us to suspect that electrons (or holes) with momenta near these singularities will 
be most important (since there are more of them). As explained in the previous section,
we obtain from these operators an $SU(4)\oplus U(1)$ algebra  of number conserving operators 
and an $SO(8)$ pairing algebra, some elements of which were used by Schulz$\>$\cite{Schulz} to
analyze pairing instabilities in generalized Hubbard models. 

The operators
\begin{equation}
\vec{\bf S^{\vphantom\dag}} = \frac{1}{2}\left(a^{\dag}_{\alpha}
\vec{\sigma^{\vphantom o}}^{\vphantom\dag}_{\!\!\alpha\beta}
a^{\vphantom\dag}_{\beta} + b^{\dag}_{\alpha}
\vec{\sigma^{\vphantom o}}^{\vphantom\dag}_{\!\!\alpha\beta}b^{\vphantom\dag}_{\beta}\right)
\hbox{\ \ ,\ \ }\vec{\bf A^{\vphantom\dag}} = \frac{1}{2}\left(a^{\dag}_{\alpha}
\vec{\sigma^{\vphantom o}}^{\vphantom\dag}_{\!\!\alpha\beta}
a^{\vphantom\dag}_{\beta}- b^{\dag}_{\alpha}
\vec{\sigma^{\vphantom o}}^{\vphantom\dag}_{\!\!\alpha\beta}b^{\vphantom\dag}_{\beta}\right)
\label{eq:SOfourN}
\end{equation}
and
\begin{equation}
Q = \frac{1}{2}\left(a^{\dag}_{\alpha}a^{\vphantom\dag}_{\alpha} 
+ b^{\dag}_{\alpha}b^{\vphantom\dag}_{\alpha} - 1\right)
\label{eq:Qdef}
\end{equation}
generate an $SO(4)\oplus U(1)$ algebra which is a symmetry of the nearest neighbor hopping Hamiltonian
(\ref{eq:Hhop}) (with $t^{\prime}=0$). The remaining number conserving operators
\begin{equation}
\tau = \frac{1}{2}\left(a^{\dag}_{\alpha}a^{\vphantom\dag}_{\alpha} 
- b^{\dag}_{\alpha}b^{\vphantom\dag}_{\alpha}\right)
\label{eq:tau}
\end{equation}
\begin{equation}
{\cal O}_{CDW} = \frac{1}{2}\left(a^{\dag}_{\alpha}b^{\vphantom\dag}_{\alpha} 
+ b^{\dag}_{\alpha}a^{\vphantom\dag}_{\alpha}\right)
\hbox{\ ,\ }
{\cal O}_{JC} = \frac{1}{2i}\left(a^{\dag}_{\alpha}b^{\vphantom\dag}_{\alpha} 
- b^{\dag}_{\alpha}a^{\vphantom\dag}_{\alpha}\right)
\label{eq:CDW}
\end{equation}
\begin{equation}
\vec{{\cal O}^{\vphantom\dag}}_{SDW} = \frac{1}{2}\left(a^{\dag}_{\alpha}
\vec{\sigma^{\vphantom o}}^{\vphantom\dag}_{\!\!\alpha\beta}b^{\vphantom\dag}_{\beta}
+ b^{\dag}_{\alpha}
\vec{\sigma^{\vphantom o}}^{\vphantom\dag}_{\!\!\alpha\beta}a^{\vphantom\dag}_{\beta}\right)
\hbox{\ ,\ }\vec{{\cal O}^{\vphantom\dag}}_{JS} = \frac{1}{2i}\left(a^{\dag}_{\alpha}
\vec{\sigma^{\vphantom o}}^{\vphantom\dag}_{\!\!\alpha\beta}b^{\vphantom\dag}_{\beta}
- b^{\dag}_{\alpha}
\vec{\sigma^{\vphantom o}}^{\vphantom\dag}_{\!\!\alpha\beta}a^{\vphantom\dag}_{\beta}\right)
\label{eq:SDW}
\end{equation}
generate instabilities (charge density wave, spin density wave, etc.).

The pairing operators
\begin{equation}
\eta \equiv \bar{a}_{\alpha}b_{\alpha}{\ ,\ }
\vec{\Pi^{\vphantom\dag}} \equiv\bar{a}_{\alpha}
\vec{\sigma^{\vphantom o}}^{\vphantom\dag}_{\!\!\alpha\beta}b_{\beta}
\label{eq:pairops}
\end{equation}
[$\bar{a}_{\alpha} \equiv {a}_{\beta}(i\sigma^y)_{\beta\alpha}$] belong to the $SO(6)$
spectrum generating algebra of the nearest neighbor Hamiltonian, and the additional 
pairing operators
\begin{equation}
\Delta_s \equiv \bar{a}_{\alpha}a_{\alpha}+\bar{b}_{\alpha}b_{\alpha}{\ \ ,\ \ }
\Delta_d \equiv \bar{a}_{\alpha}a_{\alpha}-\bar{b}_{\alpha}b_{\alpha}
\label{eq:SCops}
\end{equation}
in the $SO(8)$ pairing algebra generate instabilities corresponding to s-wave and 
d-wave superconductivity, respectively.

\section{The full SO(8) algebra}\label{sec:FullSO}

We now describe the full $SO(8)$ algebra constructed using the nesting vector {\bf Q} 
introduced above. We start with the usual charge and spin operators
\be
Q \equiv \sum_{{\bf x},\alpha}\>
\left(a^{\dag}_{{\bf x}\alpha}a^{\vphantom\dag}_{{\bf x}\alpha}-\frac{1}{2}\right)
= \sum_{{\bf k},\alpha}\>
\left(c^{\dag}_{{\bf k}\alpha}c^{\vphantom\dag}_{{\bf k}\alpha}-\frac{1}{2}\right)
\label{eq:QQdef}
\ee
\be
\vec{\bf S^{\vphantom\dag}} \equiv\frac{1}{2}\sum_{{\bf x},\alpha\beta}\>
a^{\dag}_{{\bf x}\alpha}\,\vec{\sigma^{\vphantom o}}^{\vphantom\dag}_{\!\!\alpha\beta}
a^{\vphantom\dag}_{{\bf x}\beta} = \frac{1}{2}\sum_{{\bf k},\alpha\beta}\>
c^{\dag}_{{\bf k}\alpha}\,\vec{\sigma^{\vphantom o}}^{\vphantom\dag}_{\!\!\alpha\beta}
c^{\vphantom\dag}_{{\bf k}\beta}
\label{eq:Sdef}
\ee
(we use $a^{\dag}$, $a$ to denote coordinate space operators, $c^{\dag}$, $c$ to denote
momentum space operators). 
Now divide the Brillouin zone for the square lattice into two subzones $\cal X$ and $\cal Y$,
as shown in fig.~\ref{fig:SQlattice}(b), and introduce the characteristic function (also
used by Henley$\>$\cite{Henley}){\vspace{-4pt}}
\be
\tau^{\vphantom\dag}_{\bf k}\equiv \cases{+1, &${\bf k} \in {\cal X}$\cr
-1, &${\bf k} \in {\cal Y}$}
\label{eq:tauk}
\ee
The operators ${\rm T}^{\pm}$, ${\rm T}^3$ defined by
\be
{\rm T}^{+} = \sum_{{\bf k}\in{\cal X}}\>\sum_{\alpha}\>
c^{\dag}_{{\bf k}\vphantom{+Q}\alpha}
c^{\vphantom\dag}_{{\bf k+Q},\alpha}{\ \ ,\ \ }
{\rm T}^{-} = \sum_{{\bf k}\in{\cal Y}}\>\sum_{\alpha}\>
c^{\dag}_{{\bf k}\vphantom{+Q}\alpha}
c^{\vphantom\dag}_{{\bf k+Q},\alpha}
\label{eq:Tpm}
\ee
\be
{\rm T}^3 = \frac{1}{2}\>\sum_{{\bf k},\alpha}\> 
\tau^{\vphantom\dag}_{{\bf k}\vphantom{+Q}}
c^{\dag}_{{\bf k}\vphantom{+Q}\alpha}
c^{\vphantom\dag}_{{\bf k}\vphantom{+Q}\alpha}
\label{eq:Tthree}
\ee
generate another $SU(2)\sim SO(3)$ algebra, which we call {\em isospin}; the 'flavor' here 
is simply the direction of the largest momentum component. Also
define
\be {\rm T}^1\equiv\frac{1}{2}\>\left({\rm T}^{+}+{\rm T}^{-}\right)
= \frac{1}{2}\>\sum_{{\bf k},\alpha}\>
c^{\dag}_{{\bf k}\vphantom{+Q}\alpha}
c^{\vphantom\dag}_{{\bf k+Q},\alpha}
\label{eq:Tone}
\ee
\be {\rm T}^2\equiv\frac{1}{2i}\>\left({\rm T}^{+}-{\rm T}^{-}\right)
=\frac{1}{2i}\> \sum_{{\bf k},\alpha}\>
\tau^{\vphantom\dag}_{{\bf k}\vphantom{+Q}}
c^{\dag}_{{\bf k}\vphantom{+Q}\alpha}
c^{\vphantom\dag}_{{\bf k+Q},\alpha}
\label{eq:Ttwo}
\ee
The spin $SU(2)$ and `isospin' $SU(2)$ can be enlarged to $SU(4)$ in the manner
of the supermultiplet models in nuclear theory by including the operators
\be
\vec{\bf T^{\vphantom\dag}}{\vphantom{T}}^{+} = \sum_{{\bf k}\in{\cal X}}\>\sum_{\alpha\beta}\>
c^{\dag}_{{\bf k}\vphantom{+Q}\alpha}
\vec{\sigma^{\vphantom o}}{\vphantom{T}}^{\vphantom\dag}_{\vphantom{\bf Q,}\!\!\alpha\beta}
c^{\vphantom\dag}_{{\bf k+Q},\beta} {\ \ ,\ \ }
\vec{\bf T^{\vphantom\dag}}{\vphantom{T}}^{-} = \sum_{{\bf k}\in{\cal Y}}\>\sum_{\alpha\beta}\>
c^{\dag}_{{\bf k}\vphantom{+Q}\alpha}
\vec{\sigma^{\vphantom o}}^{\vphantom\dag}_{\vphantom{\bf Q}\!\!\alpha\beta}
c^{\vphantom\dag}_{{\bf k+Q},\beta}
\label{eq:TpmV}
\ee
\be
\vec{\bf T^{\vphantom\dag}}{\vphantom{T}}^3 = \frac{1}{2}\>\sum_{{\bf k},\alpha\beta}\> 
\tau^{\vphantom\dag}_{{\bf k}\vphantom{+Q}} c^{\dag}_{{\bf k}\vphantom{+Q}\alpha}
\vec{\sigma^{\vphantom o}}^{\vphantom\dag}_{\vphantom{\bf Q}\!\!\alpha\beta}
c^{\vphantom\dag}_{{\bf k}\vphantom{+Q}\beta}
\label{eq:TthreeV}
\ee
and, analogous to (\ref{eq:Tone}) and (\ref{eq:Ttwo}),
\be
\vec{\bf T^{\vphantom\dag}}{\vphantom{T}}^1 = \frac{1}{2}
\left(\vec{\bf T^{\vphantom\dag}}{\vphantom{T}}^{+} +
\vec{\bf T^{\vphantom\dag}}{\vphantom{T}}^{-}\right){\ \ ,\ \ }
\vec{\bf T^{\vphantom\dag}}{\vphantom{T}}^2 = \frac{1}{2i}
\left(\vec{\bf T^{\vphantom\dag}}{\vphantom{T}}^{+} -
\vec{\bf T^{\vphantom\dag}}{\vphantom{T}}^{-}\right)
\ee
\vspace{-2 pt}%
The number conserving $SU(2)\oplus SU(2)\sim SO(4)$ algebra generated by the operators 
$\vec{\bf S^{\vphantom\dag}}$ and $\vec{\bf T^{\vphantom\dag}}{\vphantom{T}}^3$ commutes
with the free particle Hamiltonian $H_0$ of eq.~(\ref{eq:Hhop}) [the commuting $SU(2)$
generators are actually 
$\vec{\bf S^{\vphantom\dag}}_{\pm} = \vec{\bf S^{\vphantom\dag}} \pm
\vec{\bf T^{\vphantom\dag}}{\vphantom{T}}^3$, {\vspace{1pt}}
which are the spin generators restricted to the subzones ${\cal X}(+)$ and
${\cal Y}(-)$]. The operators $Q$, ${\rm T}^3$ also commute with $H_0$. 

We can also introduce pairing operators 
\be
\Delta = \frac{1}{2}\>\sum_{{\bf k},\alpha}\> \bar{c}_{{\bf k}\alpha}c_{{\bf k}\alpha} 
{\ \ ,\ \ }
\Delta_{\tau} = \frac{1}{2}\>\sum_{{\bf k},\alpha}\>\tau_{\bf k}
\bar{c}_{{\bf k}\alpha}c_{{\bf k}\alpha} 
\label{eq:Deltadef}
\ee
and extended pairing operators
\be
\Pi = \frac{1}{2}\>\sum_{{\bf k},\alpha}\bar{c}_{{\bf k+Q},\alpha}\>
c_{\bf{k\vphantom{+Q}}\alpha}{\ \ ,\ \ }
\vec{\Pi^{\vphantom\dag}}  = \frac{1}{2}\>\sum_{{\bf k},\alpha\beta}\>
\>\tau_{\bf k}\bar{c}_{{\bf k+Q},\alpha}
\vec{\sigma^{\vphantom o}}_{\vphantom{\bf Q}\!\!\alpha\beta}
c_{{\bf k}\vphantom{+Q}\beta}
\label{eq:Pidef}
\ee
[here $\bar{c}_{{\bf k}\alpha} \equiv (i\sigma^2)_{\alpha\beta}c_{{\bf k},\beta}$
is a `charge conjugate' operator which has the same transformation properties
as the creation operator for an antiparticle of momentum $\bf k$, spin state
$\alpha$]. The $\Delta$ and $\Pi$ operators, together with their adjoints, 
augment the $SU(4)\oplus U(1)$ algebra of number-conserving operators to form 
an $SO(8)$ pairing algebra. The operators introduced here are equivalent to those
introduced in the simplified version, as shown in table \ref{tab:SOgen}.

The largest subalgebra of the $SO(8)$ which commutes{\vspace{1pt}} with the simplest 
hopping Hamiltonian $H_0$ [eq.~(\ref{eq:Hhop}) with $t^{\prime} = 0$] is the 
$SO(6)\oplus SO(2)$ algebra generated by $Q$, $\vec{\bf S^{\vphantom\dag}}$, 
$\vec{\bf T^{\vphantom\dag}}{\vphantom{T}}^3$, $\Pi$, $\Pi^{\dag}$, 
$\vec{\Pi^{\vphantom\dag}}$, $\vec{\Pi^{\vphantom{\dag}}}\vphantom{T}^{\dag}$ and
${\rm T}^3$ [${\rm T}^3$ is the $SO(2)$ generator]. 
Note {\vspace{1pt}}that the subalgebra can also be described as 
$SU(4)\oplus U(1)$, since{\vspace{1pt}} $SU(4) \sim SO(6)$ and $U(1) \sim SO(2)$  
(this equivalence of Lie algebras is sometimes overlooked {\vspace{1pt}} in the literature). 
This $SO(6)\oplus SO(2)$ provides a starting point for analyzing collective 
instabilities.
\renewcommand{\arraystretch}{1.7}
\begin{table}[hb]
\caption{Translation table for elements of \(SO(8)\) in the simplified form  of
sec.~4 and in the full form of sec.~5.
The operators $Q$, $\protect\overrightarrow{\bf S}$ and 
$\protect\overrightarrow{\Pi}$ are common to both presentations.
\label{tab:SOgen}
}
\vspace{6pt}
\begin{center}
\footnotesize
\begin{tabular}[b]{|c|c|c|c|c|c|c|c|c|c|}
\hline
Full $SO(8)$ & ${\rm T}^1$ & ${\rm T}^2$ & ${\rm T}^3$ &
$\vec{\bf T^{\vphantom\dag}}{\vphantom{T}}^1$ & 
$\vec{\bf T^{\vphantom\dag}}{\vphantom{T}}^2$ &
$\vec{\bf T^{\vphantom\dag}}{\vphantom{T}}^3$ & 
$\Delta$ & $\Delta_{\tau}$ & $\Pi$ \\  \hline
Simple $SO(8)$ & ${\cal O}_{CDW}$ & ${\cal O}_{JC}$  & $\tau$ &
$\vec{{\cal O}^{\vphantom\dag}}_{SDW} $ &
$\vec{{\cal O}^{\vphantom\dag}}_{JS} $ &
$\vec{\bf A^{\vphantom\dag}}  $ &
$\Delta_s$ & $\Delta_d$ & $\eta$ \\
\hline
\end{tabular}
\end{center}
\end{table}

\section{Symmetries and instabilities}\label{sec:Symm}

The generators of $SO(8)$ not in the $SO(6)\oplus SO(2)$ subalgebra can be
grouped into two 6-dimensional {\em superspins}:
\be
\Sigma^{\pm} \equiv \left({\rm T}^{\pm},\vec{\bf T^{\vphantom\dag}}{\vphantom{T}}^{\pm},
\Delta \mp \Delta^{\vphantom\dag}_{\tau}, \Delta^{\dag} \pm \Delta^{\dag}_{\tau}\right)
\label{eq:supspin}
\ee
which are eigenstates of the $SO(2)$ generator ${\rm T}^3$ with eigenvalues $\pm 1$.
Generalized pairing operators of the form
\be
\Delta[\xi] \equiv \frac{1}{2}\>\sum_{{\bf k},\alpha}\>\xi({\bf k})
\bar{c}_{{\bf k}\alpha}c_{{\bf k}\alpha}
\label{eq:Deltaxi}
\ee
which generate BCS-like states can be introduced. These can also be grouped into
multiplets which transform like vectors under the $SO(6)$ group, and define instabilities 
which will compete when interactions which break the $SO(6)$ symmetry are included.

Zhang$\>$\cite{Zhang} has considered the $SO(5)$ algebra obtained from $SO(6)$ by
deleting the generators $\vec{\bf T^{\vphantom\dag}}{\vphantom{T}}^3$, $\Pi$, 
$\Pi^{\dag}$. The $SO(3)$ `isospin' algebra generated by ${\rm T}^{\pm}$, ${\rm T}^3$
commutes with Zhang's $SO(5)$, and we are left with a triplet of 5-dimensional
superspins:
\be
Z^{\pm}  \equiv \left(\vec{\bf T^{\vphantom\dag}}{\vphantom{T}}^{\pm},
\Delta \mp \Delta^{\vphantom\dag}_{\tau}, \Delta^{\dag} \pm \Delta^{\dag}_{\tau}\right)
{\ \ ,\ \ }
Z^0 \equiv  \left(\vec{\bf T^{\vphantom\dag}}{\vphantom{T}}^{3}, \Pi, \Pi^{\dag}\right)
\label{eq:Zspin}
\ee

These algebras are not exact symmetries of any `natural' two-dimensional model (though an
{\em ad hoc} model has been constructed$\>$\cite{Rabello}). The only known fully
two dimensional model with a higher symmetry is the Hubbard model, with nearest neighbor
hopping and interaction given by eq.~(\ref{eq:UHub}). This model has an 
$SO(4)\sim SU(2)\oplus SU(2)$ spectrum generating algebra$\>$\cite{Yang} consisting of 
the usual spin $SU(2)$ and a second $SU(2)$ generated by $\Pi$, $\Pi^{\dag}$ and $Q$.
There are other symmetries for two-site models of the type discussed in sec.~\ref{sec:SimpleSO},
which will be discussed elsewhere.

\section{Conclusions}\label{sec:fini}

We have described some higher symmetries and approximate symmetries which may be useful
in classifying the phase structure of quasi-two-dimensional systems such as the
cuprate high $T_c$ superconductors. Dynamical calculations are required for detailed
description -- a first step in this direction as been taken$\>$\cite{MKK}, using methods
developed originally by Balseiro and Falicov$\>$\cite{BF}.

\section*{Acknowledgments}
MTV's research is supported by the US Department of Energy under 
Grant \#DE-FG02-85ER40233.  Publication 749 of the Barnett Institute.


\section*{References}
\begin{thebibliography}{99}
\bibitem{tink}M Tinkham, {\em Introduction to Superconductivity}, 2nd ed.
(McGraw-Hill, New York, 1996)
\bibitem{KMW}V Z Kresin, H Morawitz and S A Wolf, {\em Mechanisms of Conventional
and High $T_c$ Superconductivity} (Oxford University Press, New York, 1993)
\bibitem{Rice}F C Zhang and T M Rice, \Journal{\PRB}{37}{3759}{1988}
\bibitem{PWA}P W Anderson {\em The Theory of Superconductivity in the High $T_c$ Cuprates} 
(Princeton University Press, Princeton, 1997)
\bibitem{vHs}R S Markiewicz, \Journal{\JPCS}{58}{1179}{1997}
\bibitem{Hubbard}A Montorsi (ed.), {\em The Hubbard Model: A Reprint Volume} 
(World Scientific, Singapore, 1992)
\bibitem{Schulz}H J Schulz, \Journal{\PRB}{39}{2940}{1989}
\bibitem{Henley}C Henley, \Journal{\PRL}{80}{3590}{1998}
\bibitem{BirSol}A I Solomon and J L Birman, \Journal{\JMP}{28}{1526}{1987} and
references therein.
\bibitem{MarkV}R S Markiewicz and M T Vaughn, {\JPCS} {\bf 59}, to be published;
\Journal{\PRB}{57}{14052}{1998} and unpublished.
\bibitem{Zhang}S C Zhang, {\em Science} {\bf 275}, 1089 (1997)
\bibitem{Rabello}S Rabello, H Kohno, E Demler and S C Zhang, \Journal{\PRL}{80}{3586}{1998}
\bibitem{Yang}C N Yang and S C Zhang, {\em Mod. Phys. Lett.} B {\bf 4}, 759 (1990); 
C N Yang, \Journal{\PRL}{63}{2144}{1989}
\bibitem{MKK}R S Markiewicz, C Kusko and M T Vaughn, cond-mat/9807067; 
R S Markiewicz, C Kusko and V Kidambi, cond-mat/9807068
\bibitem{BF}C A Balseiro and L M Falicov, \Journal{\PRB}{20}{4457}{1979}
\end{thebibliography}
\end{document}